\documentclass[conference]{IEEEtran}
\IEEEoverridecommandlockouts

\usepackage[hidelinks]{hyperref}
\usepackage{booktabs} \usepackage{amsmath,amssymb,amsfonts}
\usepackage{algorithm}
\usepackage{algpseudocode}
\usepackage{graphicx}
\usepackage{textcomp}
\usepackage[x11names]{xcolor}
\usepackage{ifthen}
\usepackage{xspace}
\usepackage[most]{tcolorbox}
\usetikzlibrary{shapes}
\usepackage[nohyperlinks]{acronym}
\usepackage[binary-units, per-mode=symbol, detect-weight=true, detect-family=true, group-digits=integer, group-separator={\,}]{siunitx}
\usepackage[abbreviations]{foreign}  \usepackage[keeplastbox]{flushend}

\usepackage[font={small},skip=2pt,belowskip=4pt]{caption}
\usepackage{microtype}
\usepackage[moderate]{savetrees}
\usepackage[inline]{enumitem}

\algnewcommand\algorithmicinput{\textbf{Input:}}
\algnewcommand\Input{\item[\algorithmicinput]}
\algnewcommand\algorithmicoutput{\textbf{Output:}}
\algnewcommand\Output{\item[\algorithmicoutput]}

\makeatletter
\renewcommand{\ALG@beginalgorithmic}{\small}
\makeatother

\renewcommand\subsection\subsubsection

\DeclareSIUnit\operation{ops}
\DeclareSIUnit\groupsize{\lvert \mathcal{G} \rvert}

\usepackage[style=ieee,backend=bibtex,natbib=true,sorting=none,doi=false,isbn=false,alldates=year]{biblatex}
\bibliography{anonymous}
\renewcommand*{\bibfont}{\footnotesize}

\newboolean{showcomments}
\setboolean{showcomments}{false}
\ifthenelse{\boolean{showcomments}}
{ \newcommand{\mynote}[3]{
		\fbox{\bfseries\sffamily\scriptsize#1}
		{\small$\blacktriangleright$\textsf{\emph{\color{#3}{#2}}}$\blacktriangleleft$}}}
{ \newcommand{\mynote}[3]{}}

\newcommand{\SYS}{\mbox{\textsc{A-Sky}}\xspace}
\newcommand{\accessmonitor}{\textsc{AccessControl}\xspace}
\newcommand{\writerproxy}{\textsc{WriterShield}\xspace}
\newcommand{\bbw}{\emph{BBW}\xspace}
 \acrodef{TEE}{trusted execution environment}
\acrodefplural{TEE}{trusted execution environments}
\acrodef{ANOBE}{anonymous broadcast encryption}
\acrodef{LoC}{line of code}
\acrodefplural{LoC}{lines of code}
\acrodef{TCB}{trusted computing base}
\acrodef{API}{application programming interface}
\acrodef{SGX}{Software guard extensions}
\acrodef{EPC}{enclave page cache}
\acrodef{CDF}{cumulative density function}
\acrodef{OS}{operating system}
\acrodef{CPU}{central processing unit}
\acrodef{ecall}{enclave call}
\acrodefplural{ecall}{enclave calls}
\acrodef{ocall}{outside call}
\acrodefplural{ocall}{outside calls}
\acrodef{TLS}{transport layer security}
\acrodef{REST}{representational state transfer}
\acrodef{SHA}{secure hash algorithm}
\acrodefplural{SHA}{secure hash algorithms}
\acrodef{AES}{advanced encryption standard}
\acrodef{GCM}{Galois counter mode}
\acrodef{AWS}{Amazon web services}
\acrodef{IV}{initialization vector}
\acrodef{CCA}{chosen ciphertext attack}
\acrodef{PIR}{private information retrieval}
\acrodef{GPG}{GNU privacy guard}
\acrodef{PGP}{Pretty good privacy}
\acrodef{TA}{trusted authority}
\acrodef{IP}{Internet protocol}
\acrodef{SDK}{software development kit}
\acrodef{YCSB}{Yahoo! cloud serving benchmark}
\acrodef{IoT}{Internet of things}
\acrodef{DH}{Diffie-Hellman}
\acrodef{JSON}{JavaScript object notation}
\acrodef{AE}{authenticated encryption}
\acrodef{PKE}{public key encryption}
\acrodef{JMH}{Java Microbenchmark Harness}
\acrodef{VDR}{Virtual data room}
\acrodef{RMW}{read-modify-write}
 
\newcommand\copyrighttext{  \footnotesize \textcopyright 2019 IEEE.
    Personal use of this material is permitted.
    Permission from IEEE must be obtained for all other uses,
    in any current or future media, including reprinting/republishing this
    material for advertising or promotional purposes, creating new collective
    works, for resale or redistribution to servers or
    lists, or reuse of any copyrighted component of this work in other works.
    Pre-print version. Presented in the {38th IEEE International Symposium on Reliable Distributed Systems (SRDS '19)}. For the final published version, refer to DOI \href{https://doi.org/10.1109/SRDS47363.2019.00013}{10.1109/SRDS47363.2019.00013}}
\newcommand\copyrightnotice{\begin{tikzpicture}[remember picture,overlay]
\node[anchor=south,yshift=10pt,fill=yellow!20] at (current page.south) {\fbox{\parbox{\dimexpr\textwidth-\fboxsep-\fboxrule\relax}{\copyrighttext}}};
\end{tikzpicture}}

\begin{document}

\title{Anonymous and Confidential \\File Sharing over Untrusted Clouds}

\author{\IEEEauthorblockN{Stefan Contiu\IEEEauthorrefmark{1}\IEEEauthorrefmark{3},
		S\'ebastien Vaucher\IEEEauthorrefmark{2}, 
		Rafael Pires\IEEEauthorrefmark{2},
		Marcelo Pasin\IEEEauthorrefmark{2}, 
		Pascal Felber\IEEEauthorrefmark{2} and 
		Laurent R\'eveill\`ere\IEEEauthorrefmark{1}}
	\vspace{1mm}
	\IEEEauthorblockA{
		\IEEEauthorrefmark{1}University of Bordeaux, France, \texttt{\small firstname.lastname@u-bordeaux.fr} (\IEEEauthorrefmark{3}Scille SAS, France) \\[-1pt]
		\IEEEauthorrefmark{2}University of Neuch\^atel, Switzerland, \texttt{\small firstname.lastname@unine.ch}
	}
	\vspace{-8mm}
}

\maketitle
\copyrightnotice

\thispagestyle{empty}
\pagestyle{plain}
\vspace{-4mm}
\begin{abstract}
Using public cloud services for storing and sharing confidential data requires end users to cryptographically protect both the data and the access to the data.
In some cases, the identity of end users needs to remain confidential against the cloud provider and fellow users accessing the data.
As such, the underlying cryptographic access control mechanism needs to ensure the anonymity of both data producers and consumers.

We introduce \SYS, a cryptographic access control extension capable of providing confidentiality and anonymity guarantees, all while efficiently scaling to large organizations.
\SYS leverages \emph{trusted execution environments} (TEEs) to address the impracticality of \emph{anonymous broadcast encryption} (ANOBE) schemes, achieving faster execution times and shorter ciphertexts.
The innovative design of \SYS limits the usage of the TEE to the narrow set of data producing operations, and thus optimizes the dominant data consumption actions by not requiring a TEE.
Furthermore, we propose a scalable implementation for \SYS leveraging micro-services that preserves strong security guarantees while being able to efficiently manage realistic large user bases.
Results highlight that the \SYS cryptographic scheme is 3 orders of magnitude better than state of the art ANOBE, and an end-to-end system encapsulating \SYS can elastically scale to support groups of \num{10000} users while maintaining processing costs below 1 second.

\end{abstract}

\section{Introduction}
\label{sec:introduction}

Relying on cloud services for storing content emerges as an efficient method for organizations to cut and adapt functional costs.
As cloud service providers cannot be fully trusted~\cite{bessani2014scfs}, data owners shall cryptographically protect their data before sending it to storage providers, by encrypting it with secret keys.
Furthermore, data owners grant access to well-defined groups of users to create and consume that data.
Due to lack of trust in cloud providers, cryptographic access control mechanisms are used instead to store and enforce that only valid users can access the keys and, consequently, the data.

Sometimes, not only data but also the identity of users is sensitive and has to be protected.
Consider for example military organizations that define access groups based on security clearances.
Besides protecting the shared information that is specific to a clearance level (\eg, confidential, secret and top secret), users sharing the same clearance level do not know each other.
Likewise, dispatching confidential medical programs (\eg, for HIV patients) needs to ensure that patients' privacy is guaranteed~\cite{dwyer2004health} and therefore fellow patients cannot infer their identity. 
Moreover, \acp{VDR}~\cite{iDeals} used for exchanging confidential documents during business acquisitions not only need to enforce a high-level of access control, but also to protect stakeholders' identities.

Existing research in the area of security of cloud-backed storage systems covers cryptographic access control for data confidentiality and authenticity~\cite{garrison2016practicality,popa2011enabling,contiu2018ibbe}, but not anonymity.
These systems rely on public key cryptography mapping user identities and enveloping symmetric keys that protect the actual shared content.
Differently, confidential systems focusing on group communication offer anonymity guarantees by group key exchange methods~\cite{angel2016unobservable}, requiring all active group members to be present and participate in a multi-phase protocol (\eg, Diffie-Hellman) each time a key is derived.
Such an approach is indeed suitable for instant group communication, but impractical for file sharing that generally does not require the online presence of users.
Moreover, theoretical anonymous file sharing extensions have been hypothesized~\cite{barth2006privacy,libert2012anonymous} without ever turning into functional systems.
The need for anonymous sharing of confidential content was practically addressed in an unsophisticated manner by \ac{GPG}.
The approach implemented by \ac{GPG} is to drop any public key mapping from the resulting ciphertext, and therefore keep no reference to the identity of the actual content's recipients.
The main drawback of this solution happens at decryption time, when the recipient needs to perform many asymmetric decryption trials until the portion of the ciphertext matching his private key is found (if any).
As pointed by our preliminary benchmark (\S\ref{sec:related}, Table~\ref{tab:gpg}), \ac{GPG} works well for groups of few users but quickly becomes impractical for larger ones.

As an alternative approach, \acp{TEE} such as Intel \ac{SGX}~\cite{costan2016intel} or ARM TrustZone~\cite{alves2004trustzone} have seen rapid adoption during the last few years.
Data and computations happening within such trusted environments cannot be seen from outside. 
A number of security systems profited from \ac{TEE} integration in order to achieve practical performance while targeting strict threat models~\cite{schuster2015vc3,brenner2016securekeeper}.
Envisioning \ac{TEE} usage as a building block for anonymous sharing systems is therefore natural.
However, \acp{TEE} and more specifically Intel \ac{SGX} come with side costs, notably due to transitioning latency between trusted and untrusted zones, as well as page swapping when exceeding the limited memory size of the \ac{EPC}.
In addition, one cannot rely on widespread adoption of such enabling technology.
Instead, one needs to consider the participants' heterogeneity in an anonymous sharing scheme, including various microprocessor architectures, mobile users or even \ac{IoT} devices.

In this paper, we propose an anonymous access control scheme that leverages \ac{SGX} as \ac{TEE} only for a narrow scope and deployment: enforcing anonymity during the publishing operation (\ie, upon \emph{writing}). Our scheme does not require a \ac{TEE} on the user side for performing the \emph{read} operation, nor does it require that users connect to a designated \ac{TEE} proxy.
Moreover, by leveraging \acp{TEE}, we can circumvent assumptions of state-of-the-art theoretical anonymous sharing schemes~\cite{barth2006privacy} and considerably improve the performance of cryptographic operations. 
To demonstrate the feasibility of our solution, we propose a scalable system design leveraging micro-services that can elastically scale depending on the access control and data content workloads.

Even though our work targets file sharing over untrusted cloud storages, the proposed solution can be adapted to a wider spectrum of anonymous broadcast contexts such as media streaming or peer-to-peer networks.

Our evaluation highlights that our construct is faster by 3 orders of magnitude compared to state-of-the-art \ac{ANOBE}~\cite{barth2006privacy}.
Furthermore, our end-to-end system implementation, \SYS, can adequately scale to cope with a similar number of administrative and user operations that a realistically-sized organization would experience (see \S\ref{sec:evaluation}).

In short, we propose the following original contributions:
\begin{enumerate*}[label=\emph{(\roman*)}]
	\item We define a theoretical anonymous cryptographic access control extension that relies on \acp{TEE} for a minimal subset of operations (\ie, \emph{writes} but not \emph{reads}). 
To the best of our knowledge, our approach is the first to leverage \acp{TEE} for the construction of \ac{ANOBE} primitives.
	
	\item We propose an end-to-end system design, incorporating our theoretical construct and leveraging micro-services that can scale according to the undergoing workloads.
	
	\item We implement and evaluate the system, first in isolation showing its benefits against state-of-the-art cryptographic schemes, and secondly by benchmarking its scaling capabilities and practical feasibility.
\end{enumerate*}

The paper continues by introducing the actors and adversarial threat model (\S\ref{sec:model}).
We then discuss the state of the art and open challenges (\S\ref{sec:related}),
present the design of our solution (\S\ref{sec:solution}) and its implementation (\S\ref{sec:implementation}),
evaluate our prototype within isolated micro-benchmarks and large-scale macro-benchmarks (\S\ref{sec:evaluation}),
and finally conclude (\S\ref{sec:conclusion}). 

\section{Motivation}
\label{sec:model}

We provide an overview of the assumptions and security objectives of file sharing systems that guarantee data confidentiality and user anonymity.

\subsection{Model and Use Case}

We target file sharing between \emph{users} represented by humans or software agents.
We consider that users are uniquely identified within the premises of an organization.
Users are organized into uniquely identifiable \emph{groups} by organization-specific considerations and policies.
We consider a separation between the group access control and group content management by both functional and threat factors.
Group access control represents group memberships operations and is performed by \emph{administrators}.
Administrative operations consist in adding and removing users from \emph{groups}.
Group content management represents creating and consuming files by group members.  
A user can hold one or both roles of \emph{writer} and \emph{reader} within one or multiple groups.
The remote storage is a typical cloud object storage that can store uniquely-identified large binary objects (\eg Amazon S3).

\smallskip\noindent\textbf{Exemplifying Use Case.}
\acfp{VDR}~\cite{iDeals} enable a tightly controlled exchange repository of electronic documents for company mergers and acquisitions (M\&A).
Thanks to \acp{VDR}, the seller, supporting parties assisting the seller, and acquisition bidders can confidentially exchange documents (\eg, terms, valuation) through an untrusted remote storage medium.
The seller acts as \emph{administrator} and enforces access control.
\emph{Active} user roles are constituted by \emph{writers} (the seller and supporting parties) and \emph{readers} (the bidders).
As enforced by confidentiality agreements, supporting parties operate under the umbrella of the seller, and remain unidentifiable from each other.
Similarly, the seller can conceal the identity of bidders among themselves.
As such, \emph{inner} anonymity guarantees need to be enforced within the \emph{writers} (supporting parties) and \emph{readers} (bidders) groups, while \emph{outer} anonymity needs to withstand against any actor who is not involved in the M\&A process.
Additionally, any withdrawing bidder or misbehaving supporting party can be \emph{revoked} by the seller, and therefore unable to access the document corpus.
The operation load of such scenario follows typical workloads of cloud sharing services (\eg, Dropbox, Google Drive), modeled by YCSB~\cite{cooper2010benchmarking} in our macro-benchmark evaluation (\S\ref{sec:evaluation:macro}).

\subsection{Security Objectives}
We specify four high-level security properties for confidential and anonymous file sharing systems. 

\begin{enumerate}[label=\emph{(\roman*)}]
\item \textbf{Confidentiality and Authenticity:} The secrecy of the content of shared files is exclusive for the group members.
Recipients should be able to check the integrity and provenance of shared content.
\item\textbf{Forward Secrecy:} The compromise of a group secret should not compromise past sharing sessions within the same group.
\item\textbf{Recipients Privacy:} No recipient except the group administrator should be able to infer the identities of other recipients (\ie \emph{readers}).
\item\textbf{Sender Privacy:} No recipient except the group administrator should be able to infer the sender's (\ie \emph{writer}) identity.
\end{enumerate}

\subsection{Threats}

\emph{Revoked} users and users external to the system behave arbitrarily.
They try to discover shared content and group members identities. 
To do so, they can intercept, decipher and alter exchanged messages (\ie Dolev-Yao~\cite{dolev1983security} adversarial model).

User anonymity is not only endangered by external adversaries, but also internally by considering peer group members.
As such, we consider that active users that can rightfully decrypt group content are able to launch attacks with the goal of inferring peer members identities.
To do so, they can make use of unlimited attack trials \emph{adapted} to their adversarial strategy. 
We therefore consider that the proposed solution should satisfy the strong security notion of \emph{adaptive chosen ciphertext attack} (IND-CCA2).
A solution that fulfills such guarantees also satisfies weaker security notions of non-adaptive chosen ciphertext or plaintext attacks.

The storage provider behaves in an \emph{honest-but-curious} manner.
As such, it can try to observe the incoming and outgoing data flows with the goal of discovering the actual content and the identity of the users accessing the data, all while providing service.
In order to break the confidentiality guarantee, revoked users can collude with the cloud storage to discover content created after their revocation.

Finally, our privacy model enforces the anonymity guarantee only with respect to user identities. 
We consider hiding the size of groups, how often members communicate and the size of the content that they exchange as out-of-scope.

\section{Related Work}
\label{sec:related}

This section discusses related work and open challenges in the domain of cryptographic cloud storage and access control, as well as trusted execution environments.

\smallskip\noindent\textbf{Cryptographic Cloud Storages and Access Control.}
In recent years, a number of storage and sharing system designs have been proposed for mitigating the lack of trust in cloud providers.
DepSKY~\cite{bessani2013depsky} proposes an object store interface that can be used on the client side to encrypt and redundantly store ciphertext on multiple untrusted storages.
The encryption keys are split by using a secret sharing scheme~\cite{shamir1979share} and dispersed over multiple storage systems that do not collude with each other.
SCFS~\cite{bessani2014scfs} extends the client-side encryption and cloud redundancy of DepSKY by using a trusted metadata coordination service that also encapsulates access control.

Some systems follow a different avenue by \emph{cryptographically} enforcing access control using key enveloping.
Also referred to as \emph{hybrid encryption}~\cite{garrison2016practicality}, the technique consists in encrypting data with a symmetric key that is then itself encrypted with public key encryption.
For example, CloudProof~\cite{popa2011enabling} proposes a client-side encrypted cloud storage that solves access control by using \emph{broadcast encryption}~\cite{boneh2005collusion} to envelope two keys: the first is used for decrypting (\ie, \emph{reads}) and the second one for signing (\ie, \emph{writes}).
Differently, REED~\cite{li2016rekeying} uses \emph{attribute-based encryption}~\cite{bethencourt2007ciphertext} to envelope the symmetric keys that are protecting de-duplicated content.
However, the key-enveloping technique was argued by Garrison \etal~\cite{garrison2016practicality} as impractical when target usage conditions are highly dynamic.
IBBE-SGX~\cite{contiu2018ibbe} demonstrates that the approach can be implemented within dynamic conditions when leveraging \acfp{TEE}.
Yet, none of the above constructions considers an enriched threat model for preserving both confidentiality and anonymity.

\smallskip\noindent\textbf{Confidential Messaging Systems.}
Encrypted messaging systems share a common initial phase with our cloud file sharing model, by requiring the construction of a group key that protects group communication.
Popular messaging systems (\eg, WhatsApp, Threema, Signal) use a \ac{DH} group key agreement and derivation~\cite{rosler2018more}.
Such protocols require all active participants to contribute to the creation of the group key, albeit without providing anonymity guarantees.
Pung~\cite{angel2016unobservable} uses \ac{PIR} in conjunction to a group \ac{DH} key derivation, thus achieving anonymity. 
Such a mechanism is different from our target model, in which active users do not need to participate in the creation of the group key, no matter the number of groups they belong to.

\smallskip\noindent\textbf{Pretty Good Privacy.}
In practical systems, the popular \ac{PGP}~\cite{zimmermann1995official} program, used for cryptographic protection of file or emails, addressed the anonymity criteria with a simple solution.
In anonymous mode (or \emph{hidden recipient} as called by \ac{PGP}), after performing the symmetric encryption of the content and public key encryptions of the symmetric key, all the public key mappings are dropped from the resulting ciphertext.
As such, an outside adversary cannot infer the public keys of the recipients.
At decryption time, as the recipients have no pointer to their key-envelope ciphertext fragment, they need to perform several private key decryption trials until they succeed ($\frac{n}{2}$ trials on average, where $n$ is the group size).
Table~\ref{tab:gpg} presents results of a simple benchmark of \ac{GPG} (v.\,1.4.2) in \emph{hidden recipient} mode.
One can observe that encryption and---even more notably---decryption have an impractical cost of \num{12} and \SI{60}{\second} respectively for groups of \num{1000} members.
Moreover, the inner implementation of \ac{PGP}'s \emph{hidden recipient} mode is reputed as insecure against chosen ciphertext attacks~\cite{barth2006privacy}, our targeted threat model.

\setlength{\textfloatsep}{3pt} \begin{table}
	\centering
	\caption{\label{tab:gpg} GPG operations latency in \emph{hidden recipient} mode.}
	\begin{tabular}{lSSS}
		\toprule
		Group size & {Encrypt [\si{\second}]} & {Avg. decrypt [\si{\second}]} & {Size [\si{\kilo\byte}]} \\
		\midrule
		10 & 0.13 & 0.6 & 5.3\\
		$10^2$ & 0.7 & 5.8 & 16.5\\
		$10^3$ & 12 & 60 & 129\\
		\bottomrule
	\end{tabular}
\end{table}

\smallskip\noindent\textbf{Anonymous Broadcast Encryption.}
The theoretical problem of devising a cryptographic scheme that can guarantee both confidentiality and anonymity is referred to as \emph{anonymous} (or \emph{private}) \emph{broadcast encryption} (\ac{ANOBE}).
Theoretical research literature proposes a number of such schemes, however without assessing their practicality within real systems.

The \emph{private broadcast encryption} proposed by Barth \etal~\cite{barth2006privacy} (denoted hereafter \bbw, per the authors' initials) achieves inner and outer anonymity, in addition to providing IND-CCA guarantees.
Their construction extends the public key enveloping model of \ac{PGP}, by incorporating strongly unforgeable signatures~\cite{boneh2006strongly} such that an active attacker who is member of the group cannot reuse the envelope to broadcast arbitrary messages to the group.
Moreover, to decrease the number of decryption trials, they propose the construction of publicly-known labels, unique for each member of every single encryption operation, by relying on the security assumption of \ac{DH}.
The ciphertext fragments created by the key enveloping process are therefore ordered by their label.
During decryption, after reconstructing the label, the user can seek the corresponding ciphertext fragment in logarithmic time before performing a single asymmetric decryption.
The scheme was further extended by Libert \etal~\cite{libert2012anonymous} by suggesting the use of \emph{tag-based encryption}~\cite{mackenzie2004alternatives} to hint users to their ciphertext fragment.
To the best of our knowledge, no practical system has integrated \emph{tag-based encryption} in practice.

As pointed out by our comparison benchmark (\S\ref{sec:evaluation:micro}), \bbw~\cite{barth2006privacy} can handle a key enveloping throughput of only few hundreds of users per second. 
Such a limitation requires the exploration of alternative constructions that can scale to realistic access control workloads.

\smallskip\noindent\textbf{Trusted Execution Environments.}
Recently, \acp{TEE} gathered considerable interest as an approach for solving the otherwise difficult problem of securely hosting services in the cloud, while making sure that the infrastructure provider has no knowledge of the data it handles.
Examples include performing map reduce computations~\cite{schuster2015vc3}, machine learning algorithms~\cite{ohrimenko2016oblivious} or analytics~\cite{shaon2017sgx}, while offering confidentiality guarantees to end users.

A popular choice of \ac{TEE} technology is Intel \acf{SGX}.
It defines the concept of \emph{enclave} as an isolated unit of data and code execution that cannot be accessed even by privileged code (\eg, the operating system)~\cite{costan2016intel}.
Enclaves can be \emph{attested}, that is, proving that the code that runs is the one intended, and that it is running on a genuine Intel \ac{SGX} platform.
It seems therefore natural to rely on \ac{SGX} as building block for an anonymous sharing system.
A \textit{na\"ive} approach would be to require end users to use this enabling technology and perform any access control related operations in full isolation.
We argue that this approach is impractical due to the heterogeneity of end users computing platforms, which might not be necessarily equipped with \ac{SGX} capabilities.

A different approach that makes use of \acp{TEE} is to proxy all the access control, and therefore the \emph{read} and \emph{write} operations, through a broker service that runs within enclaves.
However, we claim that given the memory and computational limitations of \ac{SGX} enclaves (\eg, \ac{TCB} size, trusted/untrusted transition latency), it is far from trivial to develop such a proxy service able to scale and sustain a high data throughput, considering dynamic access control operations~\cite{garrison2016practicality}.
While achieving scale-out by micro-services that run on top of Intel \ac{SGX} is possible in a containerized environment~\cite{arnautov2016scone}, our challenge is to define an optimal architecture for an anonymous sharing system that incorporates \ac{SGX} as \ac{TEE} with minimal performance overhead.

\section{\SYS}
\label{sec:solution}

Our solution conceptually relies on two paradigms: a cryptographic key management solution and a data delivery protocol, both designed to target an increased system performance, covering data confidentiality and user anonymity guarantees. 
We describe our solution by first having an overall look into the proposed architecture. 
We continue by detailing the design of each system operation.
Finally, we briefly discuss the security guarantees of our scheme.

\subsection{Architectural Overview}

\SYS leverages Intel \ac{SGX} as a \ac{TEE}.
In order to avoid passing all the system operations through a \ac{TEE}-enabled \emph{monitor}, we propose a design in which only data owners (\ie, \emph{writers}) are constrained to pass through such a proxy.
\emph{Readers} anonymously consume confidential content without needing to pass through the \ac{TEE}-enabled monitor, therefore not incurring in service time penalties.
The benefits of using  a monitor exclusively for write operations are manifold. 
First, the monitor acts as an outbound \ac{TA} authenticating all the content passing through.
Second, it can mask the identities of data writers. 
Third, as the monitor executes in a \ac{TEE}, traditional anonymous key management schemes~\cite{barth2006privacy,libert2012anonymous} can be modified to accommodate a new entity of trust for the key enveloping operation, therefore allowing more efficient operations. 

\begin{figure}[t]
	\centering
	\includegraphics[width=0.95\linewidth]{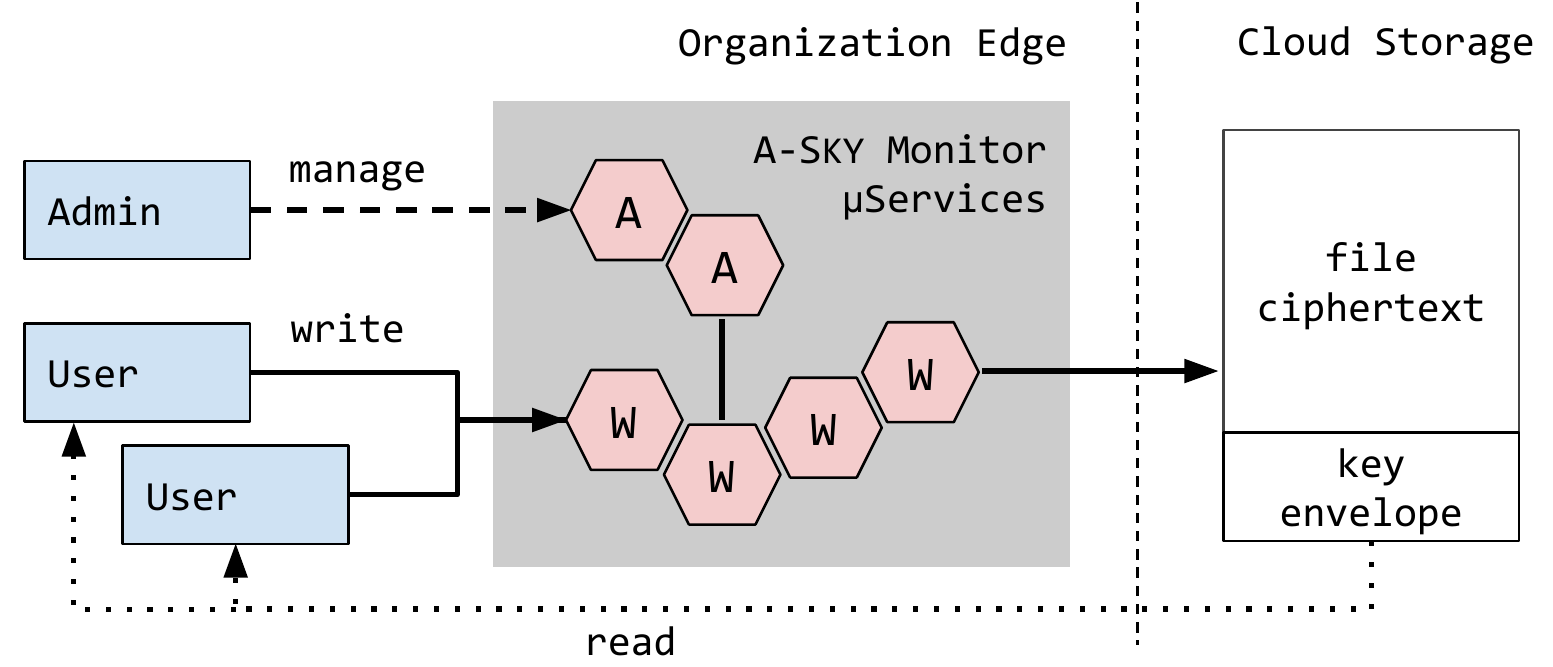}
\caption{\SYS solution overview. \SYS monitor services are \accessmonitor (\textbf{A}) and \writerproxy (\textbf{W}).}
    \label{fig:sys_arch}
\end{figure}

Fig.~\ref{fig:sys_arch} displays the overview of \SYS solution.
The \SYS monitor sits in between end-users and the cloud storage and is logically split in two roles. 
First, it provides a cryptographic mechanism for storing and enforcing access control to the data, by offering a cryptographic key management solution (the \accessmonitor service).
Second, upon successful access verification, it acts as an outgoing proxy for write operations (the \writerproxy service).
As system scalability is of paramount importance, the two logical entities (\accessmonitor and \writerproxy services) can independently adapt to undergoing load.

\smallskip\noindent\textbf{Key Management.}
The first building block of our design is a cryptographic key management solution. 
Data owners have to write content through the \ac{TEE}-enabled monitor such that only authorized readers (who are not passing through the \ac{TEE} monitor) can decrypt the data, all while having anonymity guarantees.
In traditional anonymous key sharing solutions~\cite{barth2006privacy}, a \ac{TA} performs two operations: setting up the key management system and extracting user private keys.
The operations of key enveloping together with content encryption and decryption are performed by end users. 
As such, public key cryptographic primitives are employed so that end users can cryptographically protect content for other users, whose identities are represented by public keys.
Differently, our model that leverages a \ac{TEE} as an outgoing monitor requires that the \ac{TA} does not only set up the system and extracts user keys, but also executes the key enveloping operation, which in the traditional assumptions was executed by end users.
This change of assumption therefore allows us to use a much simpler cryptographic construct to achieve the same result as traditional schemes.
Concretely, the \ac{TA} can directly make use of users' secret keys during the key-enveloping operation. 
As such, this shared secret between the \ac{TA} and end users opens the way to the use of \emph{symmetric} rather then \emph{public key} cryptography, and therefore benefit from the performance advantages of the former, \eg, hardware acceleration and smaller ciphertexts.
Second, as the traditional scheme~\cite{barth2006privacy} requires the construction of a signature key-pair per each key enveloping, under the new assumptions we can leverage the signature of the \ac{TA}. 
Moreover, the shared secret between users and \ac{TA} allows to construct efficient key de-enveloping methods that increase the performance of the decryption operation performed by end users.

\smallskip\noindent\textbf{Data Delivery Protocol.}
\SYS allows users to write the encrypted shared content through the \writerproxy service, which acts as a proxy.
The service will check with the \accessmonitor whether a user is granted the permission to write in a given group.
Being the case, it authenticates the outgoing content and does the writing itself.
We can therefore securely store the cloud storage credentials in the \writerproxy service.

\subsection{\ac{TEE} Trust Establishment}
\label{subsec:trust}

Before relying on any service of the \SYS \emph{monitor}, it is necessary to validate that the service is running on a trustworthy Intel \ac{SGX} platform, and that the instances of the \accessmonitor and \writerproxy services are genuine.
This validation phase is performed by \emph{administrators}, who are considered fully trusted (see \S\ref{sec:model}).

As such, the \ac{SGX} enclaves are required to construct a proof that incorporates the digest of the code and data inside the enclave, signed by the fused CPU private key (also known as \emph{quote}~\cite{costan2016intel}) and whose corresponding public key is retained by Intel.
The attestation packages are retrieved by the \emph{administrators}, who in turn check that the received digests are identical to known \accessmonitor or \writerproxy digests.
They then contact Intel's \emph{remote attestation service} to validate that the quote signature is indeed genuine.
Upon a successful verification, \emph{administrators} rely on the remote attestation functionality to establish a secure channel using a \ac{DH} key exchange with both the \accessmonitor and \writerproxy services~\cite{costan2016intel}.
This secure channel is used for subsequent access control operations, such as user creation, addition or removal of members from groups.
Besides, administrators are able to securely provide the cloud storage credentials to the \writerproxy service along with a long term signing key $sign_{TA}$ that is employed on all upcoming transiting content.

\subsection{Operations Design}

This subsection formally defines the operations of the \accessmonitor and \writerproxy services.

Let $E_k(p) \to c $ and $D_k(c) \to p$ define symmetric encryption and decryption algorithms using the key $k$, where $p$ is plaintext and $c$ is ciphertext.
We denote by $AE_k(p) \to (c, t)$ and $AD_k(c,t)\to \{p, \bot\}$  authenticated encryption and decryption algorithms that, besides the above symmetric primitives, can produce an authentication tag $t$ proving the integrity of the ciphertext $c$ under the key $k$.
We denote by $S_{pri}(p) \to \sigma$ and $V_{pub}(p,\sigma) \to \{true, \bot\}$ digital signature and verification schemes employing an asymmetric key-pair ($pri$ and $pub$).
Finally, $\mathcal{H}$ denotes a one way cryptographic function and $||$ denotes the literal concatenation operation.

\smallskip\noindent\textbf{\accessmonitor.} The \accessmonitor service is responsible for storing credentials, membership information and to enforce them.
Its methods are invoked by administrators through the secure channel established upon successfully performing the trust attestation process (\S\ref{subsec:trust}).

The \accessmonitor service generates user secret keys.
Given a unique user identifier~$u$, the service constructs a random secret key for the user, to whom it is sent through a \ac{TLS} channel.

The \accessmonitor service further exposes methods for group management.
Specifically, administrators can create groups, as well as add or remove users from groups.
Depending on the granted access capabilities, users can hold the roles of content \emph{reader}, \emph{writer}, or both.
The \accessmonitor service captures such capabilities within persistent dictionaries, $group^r$ and $group^w$, which store lists of users belonging to each group identifier (\eg, $group^w[g_{id}] = \{u_a, ... , u_z\}$).
Administrators are the only entities that can modify the keys and values of those two dictionaries.

\begin{algorithm}[t]
    \caption{Key enveloping (\accessmonitor)}
    \label{alg:key_envelop}
    \begin{algorithmic}[1]
        \Input user identity $u_{id}$, group identifier $g_{id}$, symmetric key $k$.
        \Output an $envelope$ ciphertext of the access control key. 
        \State $envelope \gets \emptyset$ 
        \If {$u_{id} \in group^w[g_{id}]$}
        \ForAll {users $u \in group^r[g_{id}]$}
        \State $u_{sk} \gets keys[u]$ 
        \State $(c_k, t) \gets AE_{u_{sk}}(k)$
        \State $envelope \gets envelope \cup \{(c_k, t)\}$
        \EndFor
        \EndIf
        \State\Return $envelope$
    \end{algorithmic}
\end{algorithm}

The operation of enveloping an access key for a group of anonymous members is denoted by \emph{KeyEnveloping} and is depicted in Alg.~\ref{alg:key_envelop}. 
Given the identity of the writing user, the group unique identifier, and a symmetric key $k$, the algorithm produces a ciphertext \emph{envelope} that can be anonymously de-enveloped.
The operation proceeds by first checking that the user has writing capabilities for the group (line 2).
If true, the \emph{envelope} is constructed by including the ciphertext and the authentication tag resulted from encrypting the symmetric key using the secret key of each group member (lines 3-7).

\begin{algorithm}[t]
    \caption{Proxy file (\writerproxy)}
    \label{alg:write}
    \begin{algorithmic}[1]
        \Input user identity $u_{id}$, group identifier $g_{id}$, file ciphertext $\mathcal{C}$, \accessmonitor instance $\mathcal{A}$.
        \If {$u_{id} \in \mathcal{A}.group^w[g_{id}]$}
        \State $\sigma \gets S_{sign_{TA}}(\mathcal{C})$
        \State $Upload$ to cloud : $(\mathcal{C}, \sigma)$
        \EndIf
    \end{algorithmic}
\end{algorithm}

\smallskip\noindent\textbf{\writerproxy.} 
As the \writerproxy is the sole service possessing the write credentials for the cloud provider, it constitutes a necessary hop for uploading the file.
Its main operation is $ProxyFile$ (Alg.~\ref{alg:write}).
The method verifies that the invoking user has write capabilities for the desired group (line 1).
If positive, the content is authenticated by using the long term \ac{TA} signature (line 2).
Both parts, file ciphertext and the corresponding signature, are finally uploaded to the cloud (line 3).

\begin{algorithm}[t]
    \caption{User write file to group}
    \label{alg:user_write}
    \begin{algorithmic}[1]
        \Input user identity $u_{id}$, group identifier $g_{id}$, file plaintext $P$, \accessmonitor and \writerproxy instances  $\mathcal{A}$ and $\mathcal{W}$.
        \State $fk \gets Random$ symmetric key
        \State $envelope \gets \mathcal{A}.KeyEnveloping(u_{id}, g_{id}, fk)$ \ie, Alg.~\ref{alg:key_envelop} 
        \State $cipher \gets E_{fk}(P)$
        \State $\mathcal{C} \gets envelope\ ||\ cipher$
        \State $\mathcal{W}.ProxyFile(u_{id},g_{id},\mathcal{C},\mathcal{A})$ \ie, Alg.~\ref{alg:write}
    \end{algorithmic}
\end{algorithm}

\smallskip\noindent\textbf{\textsc{User}.} The two operations performed by users are sharing a file with a group (\ie, writing) and reading a shared file. 
The user write operation leverages the \ac{TEE}-enabled monitor.
As shown in Alg.~\ref{alg:user_write}, the user first randomly creates a symmetric key (line 1) and asks the \accessmonitor service to perform an enveloping for this key  (line 2), so that it can be anonymously de-enveloped by any group member.
He then encrypts the file by using the prior generated symmetric key (line 3).
Finally, the two obtained ciphertexts---the key envelope and the file ciphertext---are concatenated (line 4) and transmitted to the \writerproxy to be uploaded to the cloud storage (line 5).

\begin{algorithm}[t]
    \caption{User read file}
    \label{alg:user_read}
    \begin{algorithmic}[1]
        \Input user secret key $u_{sk}$.
        \State $Download$ from cloud: $(\mathcal{C}, \sigma)$
        \If{$V_{pub-sign_{TA}}(\mathcal{C}, \sigma) \neq \bot$} 
        \State $envelope$, $cipher \gets split(\mathcal{C})$
        \ForAll {pairs $(k_c, t)$ in $envelope$}
        \State $fk \gets AD_{u_{sk}}(k_c, t)$
        \If {$fk \neq \bot$}
        \State $P \gets D_{fk}(cipher)$
        \State\Return $P$
        \EndIf
        \EndFor
        \EndIf
        \State\Return $\bot$
    \end{algorithmic}
\end{algorithm}

Users can read files by following the procedure of Alg.~\ref{alg:user_read}.
As previously stated, reading operations do not involve services running in a \ac{TEE}.
The first step is to download the ciphertext package from the cloud storage (line 1), that can then be validated by checking the signature (line 2) that has been appended by the \writerproxy. 
Should the signature be valid, the user then splits the package between the key envelope and the file ciphertext (line 3).
Next, the user iterates over all envelope fragments, trying to decrypt each of them by using the user secret key $u_{sk}$ (lines 4-5).
If successful, the obtained plaintext is the file encryption key, that the user can use to symmetrically decrypt the file ciphertext (lines 7-8).

\subsection{Indexing for Efficient Decryption}
\label{subsec:index}

Following the methodology of traditional \ac{ANOBE} schemes~\cite{barth2006privacy,libert2012anonymous}, we propose a method that can reduce the user decryption time by circumventing the need to perform several key decryption trials (line 4 of Alg.~\ref{alg:user_read}) by trading it off for a slight increase in key enveloping time and envelope size.
To this end, publicly known labels are constructed for each user fragment in the envelope, such that the label can be recomputed by the target recipients.
User keys are ordered by labels in the envelope, so that each key can be easily located within it and a single key decryption operation is performed.
Traditionally, the cost of building such labels was associated to performing modular exponentiation~\cite{barth2006privacy} or by using the theoretical constructs of tag-based encryption~\cite{libert2012anonymous}. 
Given the change of assumption brought by \SYS compared to traditional \ac{ANOBE}, we now have a \ac{TA} running in a \ac{TEE} performing the key enveloping.
It results that the shared secret between users and the \ac{TA} can also be used to construct efficient decryption labels. 
\SYS can therefore propose a much simpler and efficient labeling mechanism by relying on the cryptographic hash of the shared secret (\ie, the user secret key).

\begin{algorithm}[t]
    \caption{Key enveloping with \emph{efficient decryption}}
    \label{alg:key_envelop_eff}
    \begin{algorithmic}[1]
        \Input user identity $u_{id}$, group identifier $g_{id}$, symmetric key $k$.
        \Output an $envelope$ ciphertext of the access control key. 
        \State $envelope \gets \emptyset$ 
        \If {$u_{id} \in group^w[g_{id}]$}
        \State $nonce \gets Random$
        \ForAll {users $u \in group^r[g_{id}]$}
        \State $u_{sk} \gets keys[u]$ 
        \State $l_u \gets \mathcal{H}(u_{sk}\ ||\ nonce)$
        \State $(c_k, t) \gets AE_{u_{sk}}(k)$
        \State $envelope \gets envelope \cup \{(l_u, c_k, t)\}$
        \EndFor
        \State $Sort$ $envelope$ by $l$ (\ie, $label$)
        \EndIf
        \State\Return {$nonce$ $||$ $envelope$}
    \end{algorithmic}
\end{algorithm}

The efficient variant of key enveloping (Alg.~\ref{alg:key_envelop_eff}) introduces the creation of labels (line 6) as the salted hash of the user secret key. 
A random \emph{nonce} is generated for each key enveloping call to be used as a salt value, publicly included in the envelope.
The envelope fragments can therefore be sorted using the label values (line 10). 

\begin{algorithm}[t]
    \caption{User read file with \emph{efficient decryption}}
    \label{alg:user_read_eff}
    \begin{algorithmic}[1]
        \Input user secret key $u_{sk}$.
        \State $Download$ from cloud : $(\mathcal{C}, \sigma)$
        \If{$V_{pub-sign_{TA}}(\mathcal{C}, \sigma) \neq \bot$} 
        \State $nonce$, $envelope$, $cipher \gets split(\mathcal{C})$
        \State $l_u \gets \mathcal{H} (u_{sk}\ ||\ nonce)$
        \State $(k_c, t) \gets$ \emph{binary search} for $key:$ $l_u$ in $envelope$
        \If{$(k_c, t) \neq \bot$} 
        \State $fk \gets AD_{u_{sk}}(k_c, t)$
        \State $P \gets D_{fk}(cipher)$
        \State\Return $P$
        \EndIf
        \EndIf
        \State\Return $\bot$
    \end{algorithmic}
\end{algorithm}

A user read operation (Alg.~\ref{alg:user_read_eff}) requires the label reconstruction (line 4) followed by a binary search of it among the envelope fragments (line 5).
When the proper label is located, the file key can be retrieved (line 7), allowing at last the file decryption (lines 8--9).

The trade-off brought by this \emph{efficient decryption} method is therefore an overhead of $O(n\cdot \log{n})$, due to the sorting of the labels during the key enveloping operation.
The gain is reflected during decryption time, replacing $O(n)$ trials of symmetric decryption with a $O(\log{n})$ binary search and a single symmetric decryption.

\subsection{A Note on Revocation}
We argue that \SYS satisfies the \emph{lazy} revocation model~\cite{backes2006secure}, where a revoked user can continue accessing data created prior to revocation but should be unable to access any data created beyond that.
Additionally, past data becomes inaccessible upon the first succeeding write to the same resource.

The revocation is triggered by an administrator removing the user's id from the $group^r$ and $group^w$ access lists.
Later, when new content is published in that group, a new random key is derived for encrypting the content (Alg.~\ref{alg:user_write}, line 1), and a new envelope is attached to it (Alg.~\ref{alg:user_write}, line 2).
The revoked user's key will not be included in the envelope, and therefore the user will be unable to access the new group key along with the newly published content.

\section{Implementation}
\label{sec:implementation}

\subsection{\accessmonitor}
\label{sec:implementation:accessmonitor}

The \accessmonitor service is the only stateful component of \SYS.
It is responsible for generating and storing user keys, and for maintaining group membership information.
Since it deals with sensitive information, its core runs entirely within enclaves.
All external exchanges are encrypted by using \ac{TLS} connections that are terminated inside trusted environments.

We divide the \accessmonitor service into two sub-components.
The first one constitutes the entry-point for service requests.
It is developed in C++, for a total of \num{3000} \acp{LoC}.

The other one holds users and groups metadata within a replicated database.
For this purpose, we use a triple-replicated cluster of MongoDB~\cite{mongodb} servers.
MongoDB offers out-of-the-box scale-out support, and is well suited to store denormalized documents.
In order to perform queries against it from the first sub-component, we ported the official MongoDB client library~\cite{mongoc} to run inside an enclave.
Each replica of the entry-point sub-component is provisioned with the master key $M_k$ at \emph{attestation} time; its purpose is to secure the data stored in the MongoDB backend.

\begin{figure}
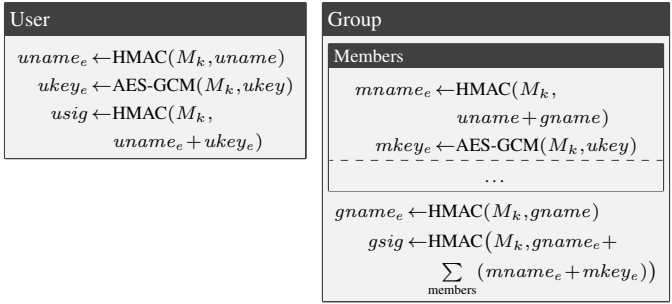

	\centering
	\centering
\tcbsetforeverylayer{
	left=0mm,
	right=0mm,
	arc=0mm,
	boxsep=0.6mm,
	top=0.3mm,
	bottom=0mm,
	boxrule=0.2mm,
}

\makeatletter
\newcommand{\changeoperator}[1]{\csletcs{#1@saved}{#1@}\csdef{#1@}{\changed@operator{#1}}}
\newcommand{\changed@operator}[1]{\mathop{\mathchoice{\textstyle\csuse{#1@saved}}
		{\csuse{#1@saved}}
		{\csuse{#1@saved}}
		{\csuse{#1@saved}}}}
\makeatother

\changeoperator{sum}

\begin{tcbraster}[raster columns=30, raster valign=top, fontupper=\scriptsize, fonttitle=\footnotesize]
	\abovedisplayskip=0pt
	\belowdisplayskip=0pt
	\begin{tcolorbox}[adjusted title=User, raster multicolumn=14]
		\begin{alignat*}{2}
			\mathit{uname_e} &\leftarrow &&\text{HMAC}\left(M_k, \mathit{uname}\right) \\
			\mathit{ukey_e} &\leftarrow &&\text{AES-GCM}\left(M_k, \mathit{ukey}\right) \\
			\mathit{usig} &\leftarrow &&\text{HMAC}(M_k, \\ & &&\mathit{uname_e}+\mathit{ukey_e})
		\end{alignat*}
	\end{tcolorbox}
	\begin{tcolorbox}[adjusted title=Group, raster multicolumn=16]
		\begin{tcolorbox}[title=Members]
			\begin{alignat*}{2}
			\mathit{mname_e} &\leftarrow &&\text{HMAC}(M_k, \\ & &&\mathit{uname} + \mathit{gname}) \\
			\mathit{mkey_e} &\leftarrow &&\text{AES-GCM}\left(M_k, \mathit{ukey}\right)
			\end{alignat*}\vspace{-4mm}
			\tcblower
			\centering\ldots
		\end{tcolorbox}\vspace{-3mm}
		\begin{alignat*}{2}
			\mathit{gname_e} &\leftarrow &&\text{HMAC}\left(M_k, \mathit{gname}\right) \\
			\mathit{gsig} &\leftarrow &&\text{HMAC}\big(M_k,  \mathit{gname_e} + \\ & &&\sum_{\text{members}} \left(\mathit{mname_e} + \mathit{mkey_e}\right)\big)
		\end{alignat*}
	\end{tcolorbox}
\end{tcbraster}
 	\caption{Data model of user and group \emph{documents} stored in MongoDB.}
	\label{fig:mongo-storage}
\end{figure}

As the storage backend runs outside of enclaves, we make sure that every piece of data that we store is either hashed using the HMAC-SHA256 construct or encrypted using \ac{AES} \ac{GCM}.
We thus guarantee that the entity that provides the MongoDB instances cannot infer any information about users or groups (barring the size of each group, which is already leaked in the \emph{envelopes}).
Fig.~\ref{fig:mongo-storage} shows how we organize data in MongoDB.
We use 2 collections, one for users and one for groups.
Each user is stored once in the users collection and once per group it is a member of.
This denormalization prevents the service provider from inferring which groups a user belongs to as the attributes of a given user are hashed or encrypted differently for each representation (\ie, we use the name of the group as salt when hashing, and different \acp{IV} when encrypting).
Each document is wholly signed using HMAC signatures to ensure its integrity.

There are two kinds of users interested in communicating with the \accessmonitor service: regular users, who need to retrieve their randomly-generated \SI{256}{\bit} private key, and administrators, who perform group access control operations.
All these interactions happen through a \ac{TLS}-encrypted REST-like protocol.
Exchanges are represented in \ac{JSON}, for which we slightly modified a C++ library~\cite{json-cpp}.
In order to terminate \ac{TLS} connections in the enclave, we use a port of OpenSSL for \ac{SGX}~\cite{openssl-sgx}.

Another duty of the \accessmonitor service is to generate key envelopes upon user requests.
An envelope contains a file key encrypted several times, once per group member.
The file key, as well as the user keys, are \SI{32}{\byte} long.
We use \ac{AES} \ac{GCM}, which generates a \emph{tag} of \SI{16}{\byte} for integrity.
Considering the addition of \SI{12}{\byte} for the \ac{IV}, each group member adds \SI{60}{\byte} to the envelope.

In order to avoid having to perform $O(n)$ decryption trials, we can index the keys within the envelope (\S\ref{subsec:index}).
First, we generate a \emph{nonce} for each envelope, that we staple to it.
Each user key is then hashed using the SHA224 algorithm, using the \emph{nonce} as a salt.
This adds \SI{28}{\byte} to the envelope for each group member.
The list of keys is sorted using the hashes as a sorting key.
As a consequence, \emph{readers} can look for their own key by doing a binary search, therefore decreasing the complexity to $O(\log{n})$ comparisons followed by one single decryption.
We evaluate the trade-offs as far as enveloping time and bandwidth usage are concerned in \S\ref{sec:evaluation}.

\subsection{\writerproxy}

The \writerproxy serves two purposes: protecting cloud storage credentials, and hiding user identities by proxying their requests to the cloud storage.
When forwarding user requests to write files, the \writerproxy checks with the \accessmonitor that the query comes from a user who has the correct permissions to write files.
User requests, including file contents, cross over the enclave boundary.
This obviously slows down transmission rates because of content re-encryption and trusted/untrusted edge transitions.
Therefore, we have also implemented a different variant where temporary access tokens are given to users, allowing them to upload their content without the aforementioned content needing to enter the \ac{TEE}.
Note that the ciphertext digest still needs to be authenticated by the signing key available in the \ac{TEE}-enabled service (necessary for IND-CCA2).
In such case, users are responsible for using appropriate proxies that can conceal the origin of the request.
One approach to hide the identities is by using peer-to-peer relay networks backed by enclaves~\cite{cyclosa2018}.
Also, it is a requirement to only communicate with the cloud storage using encrypted connections.
Even if the file data is encrypted, the metadata can leak group information to every entity listening to the network traffic.

We modeled the cloud storage component using Minio~\cite{minio}, a distributed object store that is fully compatible with the \acp{API} of Amazon S3.
As we need to perform operations against the cloud storage from within an enclave, we ported the Java version of the Minio client library to C++ so that it can run together with the \writerproxy.
These modifications amount to \num{4000} \acp{LoC} of C++, which we openly release\footnote{\url{https://github.com/rafaelppires/anonym-sharing}}.
Without accounting for external libraries, the \writerproxy consists of \num{800} \acp{LoC}.

\subsection{Client}

As part of our prototype implementation, we developed a full-featured client in \num{1200} \acp{LoC} of Kotlin.
The client can be set up to operate in all possible configurations of \SYS: keys in linear or indexed envelopes, \emph{writes} through the \writerproxy, or through a standard proxy onto a Minio or Amazon S3 storage back-end with short-lived token-based authentication.
Kotlin's full interoperability with the Java ecosystem allows us to easily integrate with the \ac{JMH}~\cite{jmh} and \ac{YCSB}~\cite{cooper2010benchmarking} frameworks that we use to perform the evaluation of \SYS (\S\ref{sec:evaluation}).

\subsection{Deployment}

All our components can be independently replicated to provide availability, fault tolerance or cope with the load.
Therefore, we have packaged our micro-services as individual containers, which we then orchestrate using an SGX-aware adaptation of Kubernetes~\cite{vaucher2018sgxaware}.
The proposed deployment considers that there exists a fast data link between the organization premises and the infrastructure where the TEE-enabled micro-services are hosted.
As such, our deployment could further benefit from \textit{edge} computing gateways sitting at the border of the organization.
Moreover, by considering that attested \ac{TEE} micro-services are self-contained with respect to the hosting environment, other deployment options arise by elastically handling \accessmonitor and \writerproxy instances between the organization edge and the user or cloud premises, if the latter two are equipped with such capabilities.
We leave these deployment options to future work.

\section{Evaluation}
\label{sec:evaluation}

We evaluate the performance and scalability of our solution by first conducting micro-benchmarks.
Then, we use the well-known \ac{YCSB}~\cite{cooper2010benchmarking} test suite to evaluate the overall system performance.

All our experiments run on a cluster of 5 SGX-enabled Dell PowerEdge R330 servers, each having an Intel Xeon E3-1270\,v6 processor and \SI{64}{\gibi\byte} of memory.
Additionally, we use 3 Dell PowerEdge R630 dual-socket servers, each equipped with 2 Intel Xeon E5-2683\,v4 CPUs and \SI{128}{\gibi\byte} of RAM.
One of the latter machines is split in 3 virtual machines to handle the roles of Kubernetes master, Minio server and benchmarking client (when a second client is needed).
SGX machines use the latest available microcode revision \texttt{0x8e}, and have the Hyper-threading feature disabled to mitigate the Foreshadow security flaw~\cite{217543}.
Communication between machines is handled by a Gigabit Ethernet network.

When error bars are shown, they represent the \SI{95}{\percent} confidence interval.

\subsection{Micro-benchmarks}
\label{sec:evaluation:micro}

\smallskip\noindent\textbf{Cryptographic Scheme Performance.}
We start the performance evaluation of \SYS by isolating and measuring the performance of the underlying cryptographic primitive.
We employ the \ac{ANOBE} scheme defined by Barth \etal~\cite{barth2006privacy} (\bbw) as a baseline.
Our implementation of \bbw uses an \emph{elliptic curve integrated encryption scheme} as the IND-CCA2 public key cryptosystem used by the original scheme.
Both cryptographic schemes key materials (\ie, keys, curve) are chosen to meet \SI{256}{\bit} of \emph{equivalent security strength}~\cite{barker2007nist}.
Moreover, we implement the efficient decryption of \bbw, as suggested in the paper by relying on the hardness of the \ac{DH} problem, however in the context of much faster \emph{elliptic curves} (ECDH).
As the content encryption is similarly implemented for the two schemes, we choose to only measure and present the key enveloping and de-enveloping performance.
We consider that the user keys are available at the time of the calls.

\begin{table}
	\centering
	\caption{\label{tab:compare} Throughput comparison (\ie, group size per second: \si{\groupsize\per\second}) of \SYS cryptographic scheme and \bbw~\cite{barth2006privacy}, isolating enveloping (Env.) and de-enveloping (Dnv.) operations, in the standard and efficient decryption ($ED$) mode.}
    \sisetup{table-format = 1.1e1, table-align-exponent = true}
    {\setlength{\tabcolsep}{1.6mm}
	\begin{tabular}{lSSSS[table-format = > 1]}
		\toprule
		& {Env. [\si{\groupsize\per\second}]} & {Dnv. [\si{\groupsize\per\second}]} & {Env.$^{ED}$ [\si{\groupsize\per\second}]} & {Dnv.$^{ED}$ [\si{\micro\second}]} \\
		\midrule
		
		\bbw & 3.3e2 & 5e3 & 3e2 & <4 \\
		
		\SYS & 1.9e6 & 2.5e6 & 1.2e6 & <4 \\
		
		\midrule
		Faster by & {\num{3.7} OoM} & {\num{2.6} OoM} & {\num{3.6} OoM} & {\emph{n/a}} \\

		\bottomrule
	\end{tabular}}
\end{table}

Table~\ref{tab:compare} shows the speed of cryptographic key enveloping and de-enveloping by reporting the number of group members handled per second.
If \bbw can envelope groups of only \num{330} members per second, \SYS can handle 3.7 orders of magnitude (OoM) more users per second.
The considerable speed difference is justified by the performance gap between public key (used by \bbw) and symmetric encryption (used by \SYS) primitives.
Likewise, a performance increase of 2.6 OoMs is observed for the de-enveloping operation.
\bbw provides an \emph{efficient decryption} mode that can achieve fast decryption times (less than \SI{4}{\micro\second} for the highest tested group size), but with a high cost of only 300 group size envelopings per second.
\SYS is able to support the same efficient decryption speed, by performing 1.2 million group size envelopings per second, a gain of 3.6 OoMs compared to \bbw.
Furthermore, as explained in \S\ref{sec:implementation}, \SYS produces a ciphertext of \SI{60}{\byte} and \SI{88}{\byte} respectively for the standard and efficient decryption modes, per each group member, compared to \SI{126}{\byte} and \SI{154}{\byte} bytes per member for \bbw.

\begin{figure}
	\centering
	\includegraphics[width=\linewidth]{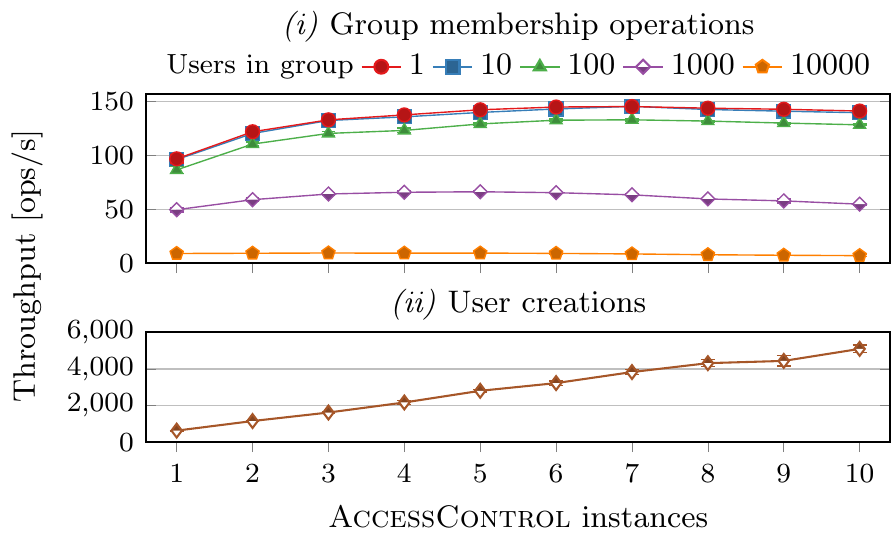}
	\caption{Throughput achieved by \accessmonitor: \emph{(i)}~adding or revoking users to/from groups of various sizes, and \emph{(ii)}~creating users.}
	\label{fig:admin}
\end{figure}

\smallskip\noindent\textbf{Scalability.}
We further evaluate the throughput of operations performed by administrators when varying the number of \accessmonitor instances.
Requests are distributed among the instances of \accessmonitor by exposing a \emph{service} in Kubernetes.
Fig.~\ref{fig:admin} shows our results.
The scalability of adding a user to a group or revoking its rights is limited, as these operations require to perform one \iac{RMW} cycle to check and update the signature of the group \emph{document}.
The larger the group, the more the operation takes time as each signature encompasses every user within the group.
This effect could be mitigated by, \eg, batching multiple operations on a given group together.
On the other hand, the operation that creates users scales linearly with the number of \accessmonitor instances, allowing more than \num{5000} user creations per second with \num{10} instances.

\begin{figure}
	\centering
	\includegraphics[width=\linewidth]{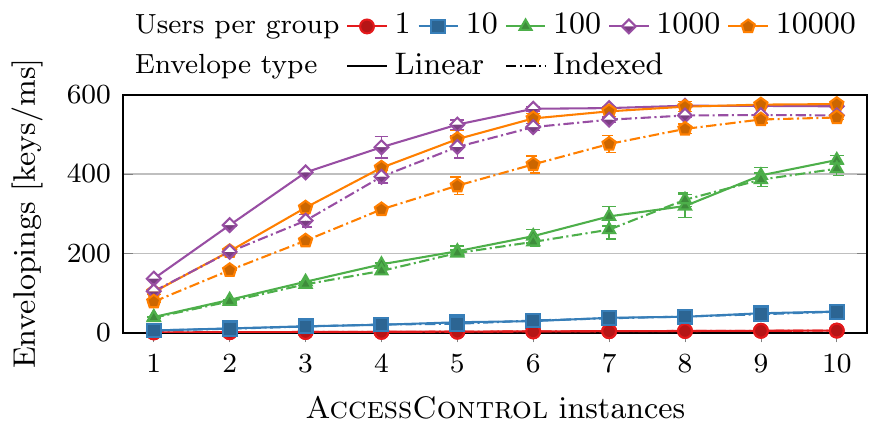}
	\caption{Throughput of enveloping a message for groups of various sizes with varying instances of the \accessmonitor micro-service.}
	\label{fig:envelope-tput}
\end{figure}

Next, we evaluated the number of keys that can be included in an envelope per unit of time, also when varying the number of instances of the \accessmonitor service.
A close-to-linear trend can be observed according to number of instances in Fig.~\ref{fig:envelope-tput}, showing that this operation also benefits from horizontal scalability.
With groups of \numrange{1000}{10000} members, the throughput ceases to increase with more than \num{7} instances as the MongoDB backend becomes a bottleneck. 
For smaller groups, the performance is diminished due to the overhead associated with each request (\eg, network connection, REST request, enclave transitions, \etc), although increasing the number of \accessmonitor instances provides greater benefits.
Additionally, we ran the same experiment with the indexing feature turned on.
For groups of \num{10000} users, the throughput is reduced by \SIrange{6}{26}{\percent}, having a marginal impact on smaller groups where the performance mostly depends on fixed costs.

\begin{figure}
	\centering
	\includegraphics[width=\linewidth]{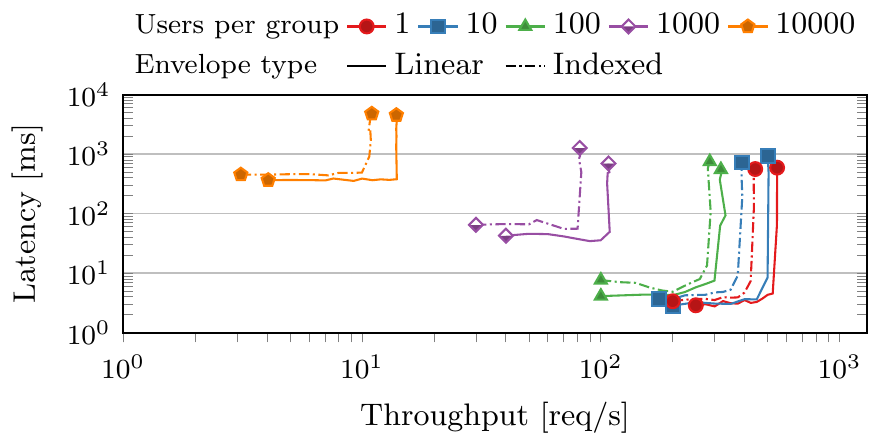}
	\caption{Throughput \emph{vs.} latency plot of enveloping a message for groups of various sizes.}
	\label{fig:envelope-tput-lat}
\end{figure}

We also evaluated the latency of the enveloping operation by increasing the throughput until saturation, again with indexing turned off and on.
Looking at Fig.~\ref{fig:envelope-tput-lat}, we notice that for groups which are larger than \num{100} users, latency increases linearly according to the group size, while the saturation throughput decreases linearly.

\begin{figure}
	\centering
	\includegraphics[width=\linewidth]{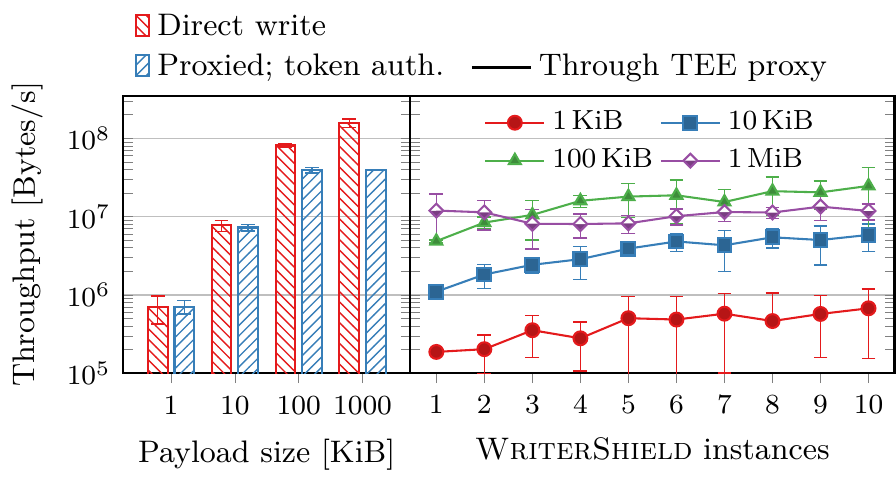}
	\caption{Throughput of writing data to the cloud storage in different ways: directly (baseline), through a TCP proxy using a temporary token for authentication, and through varying number of in-enclave \writerproxy instances.}
	\label{fig:writer-proxy}
\end{figure}

To evaluate the performance of the \writerproxy, we conduct two experiments.
In the first one, data written to the cloud is proxied through the \ac{TEE}.
In the second one, the \writerproxy is only used as a facilitator to obtain temporary access tokens for the cloud storage, with write operations being proxied through an NGINX server in TCP reverse-proxy mode.
In order to establish a baseline, we also wrote the data directly to the cloud storage service, without any intermediary.
Results are shown in Fig.~\ref{fig:writer-proxy}.
Looking at the bar plot on the left-hand side, we notice that, for files of \SI{1}{\kilo\byte} and \SI{10}{\kilo\byte} the difference in performance is negligible, whereas bigger files cause more performance degradation when using the token feature.
When the \writerproxy is used to forward data instead (right-hand side), we see that the throughput increases with the number of service instances until it seems to plateau at about the same values as with the tokenized variant.
For files of \SI{1}{\mega\byte}, adding \writerproxy instances shows no benefit.
This effect happens due to the saturation of enclave resources acting as a TLS bridge between clients and the cloud storage server.
Overall, using tokens would be the most efficient approach, although in this case the client would be responsible for using adequate proxies in order to hide its identity from the cloud storage.

\vskip 6pt
\subsection{Macro-benchmarks}
\label{sec:evaluation:macro}

\setlength{\textfloatsep}{12pt} \begin{figure}
	\centering
	\includegraphics[width=\linewidth]{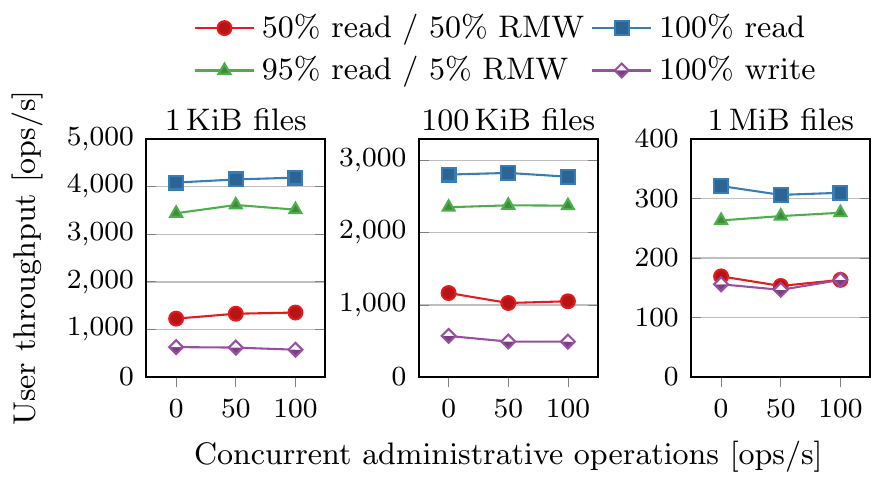}
	\caption{User throughput observed by our YCSB-based macro-benchmark, with various file access patterns, varying file sizes, and addition of simultaneous administrative operations.}
	\label{fig:macrobenchmark}
\end{figure}

We use \ac{YCSB}~\cite{cooper2010benchmarking} to observe the behavior of \SYS under different usage conditions that are specific to data serving systems.
We implemented an interface layer to link the benchmarking tool to \SYS.
As our system is not capable of direct-access writes, \emph{update} operations are replaced by \ac{RMW} operations.
In order to capture usage conditions, we run \ac{YCSB} workloads \emph{A} (update heavy), \emph{B} (read heavy) and \emph{C} (read only), to which we add an \emph{insert}-only workload.
We consider files of 3 different sizes from \SI{1}{\kibi\byte} to \SI{1}{\mebi\byte}.
We simulate \num{100000} operations across 64 concurrent users and report upon the user operation throughput.
At times, we add a second simultaneously-running instance of \ac{YCSB} that simulates 8 administrators doing group membership operations.
The administrative operations are equally distributed between adding a user to a group and revoking one, so that the size of the user database stays more-or-less constant.

Fig.~\ref{fig:macrobenchmark} shows the results of our experiment.
One can notice that the user throughput is not influenced by concurrent administrative operations, as each type of operation involves separate components of our architecture.
For small files of \SI{1}{\kibi\byte}, an increasing proportion of writes causes a degradation in performance from \SI{4100}{\operation\per\second} for read only to \SI{628}{\operation\per\second} for write-only workloads.
With larger \SI{1}{\mebi\byte} files, the difference is more nuanced, with a throughput of \SI{320}{\operation\per\second} for the read-only workload compared to \SI{155}{\operation\per\second} for the write-only workload.
Therefore, the fixed costs are largely dominant when writing small files (\eg, enveloping the file key), but are increasingly amortized for larger file sizes.
We can also observe that the throughput in \si{\byte\per\second} (\ie, multiplying the result in \si{\operation\per\second} to the file size) is largely superior for larger files, as we have already noticed in Fig.~\ref{fig:writer-proxy}.
In a nutshell, we retain that the end-user experience offered by \SYS is not influenced by concurrent administrative operations, and that the overhead of the additional operations required for writing become smaller for larger files.

\section{Conclusion}
\label{sec:conclusion}

\acresetall

We introduced \SYS, an end-to-end system that guarantees anonymity and confidentiality of shared content (\eg, files).
\SYS leverages \acp{TEE} exclusively for the content sharing operation, while \ac{TEE} capabilities are not required for end users consuming the shared content.
We have introduced a novel \ac{ANOBE} scheme that exploits additional assumptions about the availability of a \ac{TEE} compared to state-of-the-art schemes, in order to achieve fast and practical performance for its operations.
We incorporated the novel cryptographic construction into a scalable system design that leverages micro-services to elastically scale per the undergoing access control and data sharing workloads.
Results indicate that our cryptographic scheme is faster than the state-of-the-art \ac{ANOBE} schemes by 3 orders of magnitude.
An end-to-end system that utilizes our scheme can serve groups of \num{10000} users with a throughput of \num{100000} key derivations per second per service instance.
\vskip 5mm 
\vspace{-10pt}
\section*{Acknowledgment}

The research leading to these results has received funding from the French Directorate General of Armaments (DGA) under contract RAPID-172906010.

\printbibliography

\clearpage

\pagebreak
\section*{Appendix}
\smallskip\noindent\textbf{Security analysis.}
We discuss the security guarantees of \SYS and provide a brief intuition for the formalism of a reductionist security proof. 
We hypothesize that \SYS achieves indistinguishability with respect to adaptively-chosen ciphertexts (\ie IND-CCA2) according to our targeted threat model (\S\ref{sec:model}).
Within IND-CCA2 security, the adversary can be an active member of the group and therefore can rightfully decrypt group messages.
Such an attacker is allowed to try as many additional group encryptions (\ie, key envelopings) of arbitrarily constructed groups, without being able to infer if the resulted ciphertexts (\ie, \emph{envelopes}) are pointing to the same group members.
Note that proving \SYS as IND-CCA2 implicitly assures security guarantees against \emph{non} adaptive chosen ciphertext (IND-CCA) and plaintext (IND-CPA) attacks. 
Differently than IND-CCA2, IND-CCA assumes that the adversary is given only one chance to try a set of group encryptions.
Within IND-CPA, also known as \emph{semantic security}, the adversary is a passive group member that only observes and does not have the ciphertext choice capability.
Intuitively, the security guarantees extend to adversaries that are not members of a group.

Before laying out the security proof sketch, we recall the two pillars of \SYS: \acf{AE} and \acfp{TEE}.
\ac{AE} primitives are considered secure in the adaptive chosen ciphertext attack when employing the \emph{encrypt-then-mac} composition method~\cite{bellare2000authenticated}.
Such a guarantee forces to choose a specific \ac{AE} mode for \ac{AES}, as described in \S\ref{sec:implementation}.
On the other side, \acp{TEE} have been used in the composition of functional encryption cryptographic primitives shown to achieve IND-CCA2 guarantees~\cite{fisch2017iron}.
In the following, we retain the formalism of Fisch \etal~\cite{fisch2017iron} that abstracts \acp{TEE} as a \emph{secure hardware scheme}.

\textbf{Theorem 1}. Assuming that \ac{AE} is IND-CCA2 and a \ac{TEE} is a \emph{secure hardware scheme}, then \SYS is IND-CCA2.	

\textbf{Proof (intuition).}
We provide the sketch of a reductionist method that lays the frame for a formal proof.
\SYS can be seen as a reduction of the anonymous broadcast encryption scheme of Barth \etal~\cite{barth2006privacy} (\bbw), by considering two arguments:
(1)~\ac{AE} in conjunction with a \emph{secure hardware scheme} replaces the \ac{PKE} scheme, and (2)~\emph{secure hardware scheme} signatures replace the strongly unforgeable signature.
As \bbw has been proved IND-CCA2 secure by Libert \etal (Theorem~1 of~\cite{libert2012anonymous}), relying on the two aforementioned replacements, one can construct identical adversary-challenger game steps (Def.~2 in~\cite{libert2012anonymous}) and employ a similar sequence of experiments (Appendix A in~\cite{libert2012anonymous}) that can prove that \SYS is immune to chosen ciphertext attacks.
As such, in the formal language of computational security proofs, \SYS security relies on the assumption that an attacker would be unable to employ a polynomial time Turing machine for breaking the computational hardness of authenticated encryption and the robustness of \acp{TEE}.

 \end{document}